# QoE Support for Multi-Layered Multimedia Applications

Hengky Susanto, ByungGuk Kim, and Benyuan Liu

Department of computer science
University of Massachusetts at Lowell
*{hsusanto,bliu,kim}@cs.uml.edu*

**Abstract – Congestion control protocol and bandwidth allocation problems are often formulated into Network Utility Maximization (NUM) framework. Existing solutions for NUM generally focus on single-layered applications. As applications such as video streaming grow in importance and popularity, addressing user utility function for these multi-layered multimedia applications in NUM formulation becomes vital. In this paper, we propose a new multi-layered user utility model that leverages on studies of human visual perception and quality of experience (QoE) from the fields of computer graphics and human computer interaction (HCI). Using this new utility model to investigate network activities, we demonstrate that solving NUM with multi-layered utility is intractable, and that rate allocation and network pricing may oscillate due to user behavior specific to multi-layered applications. To address this, we propose a new approach for admission control to ensure quality of service (QoS) and quality of experience (QoE).**

## I. INTRODUCTION

The explosive growth and popularity of streaming applications such as video content has put immense pressure on network requirements and performance. The study of managing network congestion is often formulated into Network Utility Maximization (NUM) framework [1,2]. The objective is to allocate bandwidth that maximizes total user utility, subject to network capacity constraints. Existing approaches often model user satisfaction according to the network traffic elasticity of single-layered applications, as discussed in these literatures [1,2,3,4,5,13]. However, such model is insufficient to capture the characteristics of applications with multi layers of quality, such as video streaming. The characteristics of multi-layer applications allow users to be adaptive with their demand for QoE and QoS, which is not possible in single layered applications. In this paper, we propose a design for user utility function for multi-layered applications that incorporates studies from computer graphics, and we investigate how it may impact network activities using NUM framework, particularly for video streaming applications.

To better address the particularity of user utility function for multimedia applications, we design our utility function to reflect the characteristics of multi-layered encoding schemes, which is often used in multimedia applications. That is, the level of user utility is not just measured by the ability to meet the minimum required QoS, but also by the varying degrees of qualities associated with each encoding layer. The proposed user utility is guided by studies of human visual perceptions in the fields of computer graphics to ensure the accuracy of modeling the QoE aspect. In essence, we derive three important insights: the unique adaptive nature of multimedia applications, users are willing to tolerate some level of disruption for the sake of better image quality, and human ability to detect improvement in image quality is not infinite, i.e. it reaches a certain point where human eyes are no longer able to detect further improvement in image quality. These insights contribute to our *staircase*-shaped user utility function that follows the law of diminishing returns. This utility function also illustrates that users may have different levels of QoE, which means users can be adaptive with their demand to achieve the desired QoE. Hence, we propose a model to encapsulate this user's demand adaptability to achieve the desired QoE. Furthermore, the model also considers the impact of user's willingness to pay (budget) for the desired QoE. We then use these models to investigate network activities by incorporating the multi-layered utility function into NUM framework. Our results show that the algorithm used to resolve NUM may not converge when users are actively seeking to meet their desired QoE, resulting in oscillation in bandwidth allocation and network price. Furthermore, the oscillation also causes frequent quality adaptation at the video application level and creates a visual *flickering effect*, which degrade QoE and users find the effect annoying [27].Moreover, our results also show that the oscillation ripples to different parts of the network, and cause bandwidth allocation to other users to oscillate too. This makes solving NUM problem with multi-layered user utility to become intractable, therefore, it may not have an optimal solution. To resolve this problem, we propose greedy based network admission control to ensure that acceptable QoE and QoS are achieved.

This paper is organized as follows. We begin by discussing previous related work. Following this, we present our major contributions: the proposed multi-layered user utility model and a discussion on how the new model impacts network activities in section III and IV respectively. We introduce admission control in section V. The simulation results and discussion are presented and discussed in section VI, followed by concluding remarks.

## II. RELATED WORK

A number of proposals address NUM problems for multi-layered applications by incorporating rate video distortion minimization into user utility function. In [11], the authors' objective is to allocate bandwidth that minimizes video rate distortion, where lower rate of distortion increases higher user satisfaction. Likewise, in [16], peak signal noise ratio (PNSR) is incorporated into user utility function in their NUM formulation. In [15], the authors develop network coding based utility maximization model by integrating a taxation-based incentive mechanism to address how many layers each user should receive and how to deliver them. Fundamentally, these proposals follow the convention of single layer utility function. That is, the layer used in the solution is predetermined and not adaptive, even though it is designed to support multi-layered applications.

Others propose to adopt a staircase shape or stepwise utility function. In [11], a staircase shape rate distortion- based utility function is incorporated into NUM frameworks in multicast network environment. Different weight factors are assigned to utility function, corresponding to the quality of the layer. The authors of [12] suggest an approach that connects different points in the staircase to produce a concave shape, which in turn provides some approximate rate change. However, it is still difficult to fine tune the variables to achieve a tight approximation. Similarly, in [28], staircase shape utility function is approximated using the log function. The idea is that every level in the staircase is represented by a concave shape utility function and the authors assume the optimal solution falls within one of the steps of the staircase. In both [12,28], the encoding layer is determined after bandwidth is allocated and user utility is measured according to the allocated bandwidth. In contrast, in our model, user utility is measured according to the quality of experience user has with the image quality, rather than plainly based on the amount of bandwidth allocated. In addition, our model considers a critical characteristic of multi-layered applications – the adaptive nature of such applications which adjusts its quality according to the traffic load in order to deliver the best QoE possible to users.

## III. MULTI-LAYERED USER UTILITY FUNCTION

### A. *Foundations*

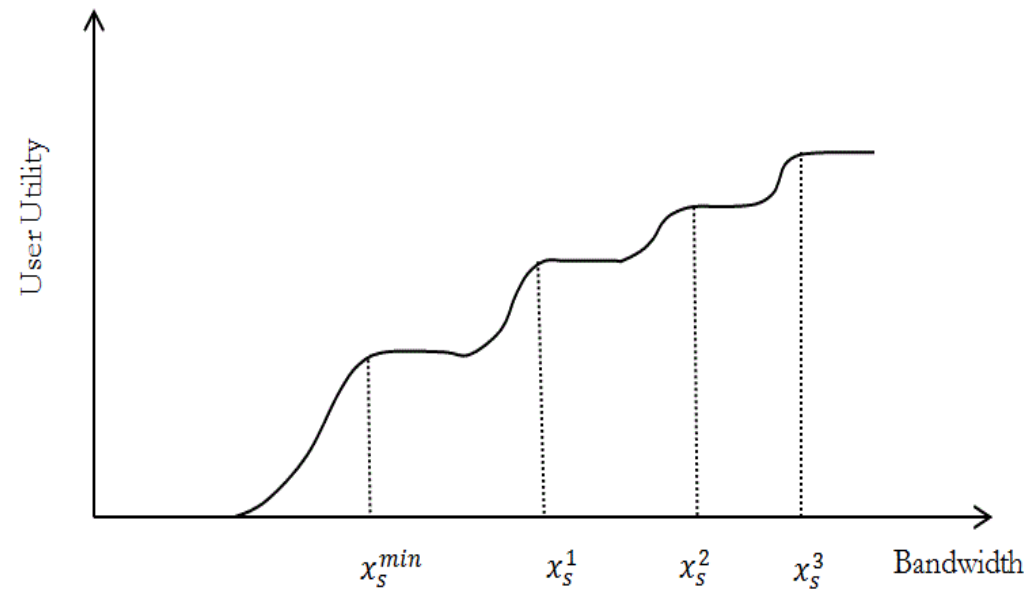

*Fig. 1: Staircase user utility function for multi layered multimedia applications.*

Multi-layered user utility function is modeled according to layered encoding scheme to reflect how multimedia traffic is managed at the network and at the application level. This means video information is divided into several encoding layers in order to minimize the amount of bandwidth used, with lower layers containing low resolution information and higher layers finer quality ones [10]. This strategy allows the video provider to provide a range in video qualities. This strategy is also known as hierarchical coding. Network forwards only the number of layers that the physical link can support and drops layers selectively at the bottleneck link. Thus, the hierarchical video encoding provides the foundation of staircase-like user utility function illustrated in figure 1.

Then we incorporate knowledge of how human eyes perceive and evaluate image qualities in our utility model because QoE is significantly affected by our visual ability. A study in [21] shows as video is encoded with more layers, discerning differences in quality between images at similar levels of quality becomes progressively difficult as the quality improves. This is because human visual capacity is less sensitive to high frequency image details and more sensitive to lower frequency [10,18,19]. Thus, we assert that human visual perception actually follows the law of diminishing returns, i.e. the benefits of offering higher quality diminishes as quality improves. Our user utility function incorporates these factors in the QoE, demonstrated by the progressively decreasing heights of the steps in the staircase-shaped function. In addition, supporting studies show that degradation in video is even less noticeable when there is a high degree of motion in the imagery [20], such as in action movies or sports videos, and especially when images are combined with good audio quality [21].

### B. *User Utility Function*

By incorporating layered encoding scheme and limitations of human's visual perception, our user utility function for multi-layered applications is modeled as follows. Since the staircase in the user utility model obeys the law of diminishing returns, the height of the steps progressively flattens toward the maximum quality, similar to an upward moving escalator. Let Y be a set of video encoding layers, $Y = \{1, 2, 3, \ldots, |Y|\}$, and $y \in Y$ when $y = 1$, $y$ is the lowest encoding layer; and when $y = |Y|$, $y$ is the highest encoding layer with the maximum quality. Let $x_s$ be the amount of bandwidth allocated to user $s$ and $x_s^y$ denotes the amount of bandwidth needed by user $s$ to support the application quality at layer $y_i$, such that $x_s \geq x_s^y$, which means network allocates at least $x_s^y$ to support the quality at layer $y$. Furthermore, let $U_{bw}(x_s^y)$ denote the user utility function of user $s$ for the amount of bandwidth $x_s$ at layer $y$, the relationship between user utility of a given layer can be illustrated as

$$\lim_{y \to \infty} \; U_{bw}(x_s^y) - U_{bw}(x_s^{y-1}) = 0.$$

Subsequently, by incorporating multi layers of encoding into the user utility function, function $U_{bw}(x_s)$ is defined as follows:

$$U_{bw}(x_s^y) = \sum_{i=0}^{|Y|} \left( U_s^y(x_s^y) \;\; \psi_U(x_s^y) \right), \qquad (1)$$

where $U_s^y(x_s^y)$ denotes user utility function given the required amount of bandwidth $x_s^y$ to support the quality at layer $y$ and $\psi_U(x_s^y)$ denotes the decay factor function at each layer and a good candidate function is $\psi_U(x_s^y) = e^{-\omega_y \, x_s^y}$, where $\omega_y$ is a positive variable used for normalization. Moreover, to ensure the minimum bandwidth required for the lowest encoding layer is met, condition $x_s \geq x_s^1$ must be met. Thus, $U_s^y(x_s) = \frac{1}{(1+e^{-a x_s^y})}$. Next, we introduce $U_{cost}(x_s)$, the user utility function that measures user satisfaction over the cost of the service relative to user's willingness to spend $m_s$. Thus, $U_{cost}(x_s) = 1 - \frac{x_s \lambda_s}{m_s}$. Therefore, user utility function for multi-layered applications has the following properties:

1. $U_{bw}(x_s^y)$, $U_{cost}(x_s) \geq 0$, $\forall x_s, 0 \leq x_s^y \leq x_s \leq C_l$, and $U_{bw}^y(0) = 0$, $U_{cost}(0) = 0$, $\forall x_s, s$, where $C_l$ is the capacity of link $l$.
2. $U_{bw}(x_s^y)$ is twice differentiable.
3. $U_{bw}(x_s^y)$ is *staircase-shape* like $\forall x_s, 0 \leq x_s^y \leq x_s$, $y = \max(Y | x_s^y \leq x_s)$, and $U_{bw}(x_s^y) \leq U_{bw}(x_s)$.
4. $U_{bw}(x_s^y)$ follows the law of diminishing returns.
5. $U_{cost}(x_s)$ is linear.
6. $\frac{\delta U_{bw}(x_s^y)}{\delta x_s^y} < \infty$, for all $0 \leq x_s^y \leq x_s \leq c$.
7. $\lim_{x_s \to 0} \frac{\delta U_{bw}(x_s^y)}{\delta x_s^y} < \infty, \forall s,\ s \in S$ and for $x_s^y \leq x_s$.
8. For $\forall y, s,\ y < y+1$, $x_s^y < x_s^{y+1}$, for $s \in S$ and $y \in Y$.

Then, the user utility function for multi-layered application is defined as follows.

$$U_s(x_s, x_s^y) = U_{bw}(x_s) + U_{cost}(x_s), \tag{2}$$

such that each user maximizes his/her own utility function by solving

$$maximize\ \ U_s(x_s, x_s^y) \tag{3.a}$$

$$s.t \quad y < y+1,\ \forall y \in Y, \tag{3.b}$$

$$\text{over } x_s \geq x_s^y \geq 0. \tag{3.c}$$

The constraint (3.b) describes that a higher layer is only considered when lower layers are delivered, that is, layer $y$ is attained if and only if all layers lower than $y$ are attained. How user solves the problem (3) above will be discussed in the following section.

### *C. System Setup*

Consider a network with a set of links $L$, and a set of link capacities $C$ over the links. Given a utility function $U_s(x_s)$ of user $s$ with the allocated bandwidth of $x_s$, the NUM formulation becomes

$$max \sum_{s \in S} U_s(x_s, x_s^y), \tag{4.a}$$

$$s.t. \quad \sum_{s \in l} x_s \leq C_l, \forall l \in L \tag{4.b}$$

$$y < y+1,\ \forall y \in Y, \tag{4.c}$$

$$over\ \ 0 \leq x_s^{min} \leq x_s,\ \forall s \in S \tag{4.d}$$

where $S$ denotes a set of users, $x_s^{min}$ denotes the minimum bandwidth requirement of user $s$, and $s \in l$ means user who is transmitting data through link $l$. The constraint (4.c) is necessary to ensure that the higher layer depends on lower layer.

Typically, the common solution to NUM is *subgradient based method* [3], and the dual problem $D$ to the primal problem of (4) is constructed as follows. $L(x, \lambda) = \sum_{s \epsilon S} U_s(x_s, x_s^y) - \sum_{s \epsilon S} \lambda_s x_s + \sum_{l \epsilon L} \lambda_l C_l$, where $L(x, \lambda)$ is the Lagrangian form and $\lambda$ is known as a set of Lagrangian multipliers $\lambda_l$, which is often interpreted as the link cost and $\lambda_s = \sum_{l \epsilon r_s} \lambda_l$. The dual problem $D$ is then defined as

$$\min D(\lambda), \quad s.t\ \lambda \geq \bar{0},$$

where the dual function $D(\lambda) = \max_{\bar{0} \leq x \leq x^{max}} L(x, \lambda)$. User decides the transmission rate $x_s(\lambda_s)$ at price $\lambda_s$ by solving

$$x_s(\lambda_s) = \underset{0 \leq x \leq x^{max}}{\arg\max} \left(U_s(x_s, x_s^y)\right), \tag{5}$$

where $x_s(\lambda_s)$ denotes bandwidth allocation at price $\lambda_s$. A subgradient projection method is used in [3]. Thus, the network on each link $l$ updates $\lambda_l$ on that link:

$$\lambda_l^{t+1} = \left[\lambda_l^t - \sigma^t \left(C_l - \sum_{s \in l} x_s\right)\right]^+, \tag{6}$$

where $C_l - \sum_{s \in l} x_s$ is a subgradient of problem $D$ for $\lambda_l^t \geq 0$. The time $t$ denotes the iteration index, for $0 \leq t$, and $\sigma^{(t)}$ denotes the step size to control the tradeoff between a convergence guarantee and the convergence speed, such that $\sigma_l^{(t)} \to 0$, as $t \to \infty$ and $\sum_{t=1}^{\infty} \sigma_l^{(t)} = \infty$. Price paid by users is $\lambda_s = \sum_{l \in s} \lambda_l$, where $l \in s$ means the link that is used by $s$ to transmit data.

## IV. ADAPTIVE USER DEMAND

The authors of [26] claim that, regardless of the algorithm, deciding the proper quality in dynamic network is essentially difficult because of a lack of information and transparency between OSI layers. Often, user experience at the application layer relies on users to decide and adapt to reach their desired quality. For that reason, we investigate how multi-layered user utility function mirrors user QoE and decision for bandwidth demand, and how these two factors impact network activities.

### *A. Adaptive Demand*

The staircase like function allows user to be adaptive with their requirement to achieve the desired QoE, as long as the minimum demand is met. However, network does not distinguish between user utility of single or multi-layered applications. The implication is, the bandwidth allocation which is sufficient for users of single-layer applications may not be sufficient users of multi-layered applications. This is because users using multi-layered applications may continue to demand more bandwidths to improve their experience. This is noted in HCI studies in [27] where viewers prefer jerking (or less smooth) video with better image quality over smooth but poor visual quality where objects in the video are not recognizable. Low tolerance for poor image quality may motivate users to demand for more bandwidths for better image quality. In other words, due to the dynamic nature of user demand, providing the minimum bandwidth requirement does not automatically maximize the aggregated user satisfaction level because there are varying levels of user satisfaction in a multi-layered environment. Conversely, a user may stop demanding for additional bandwidth as they consider cost of service, especially when a user is already experiencing a sufficiently high quality of service, when further improvement in image quality cannot be appreciated because our eyes are not able to detect the quality difference.

*B. User's Desire for Better Quality*

In order to investigate the impact of the adaptive nature of multi-layered utility on network activities, we first provide a model that encapsulates user's motivation to stay put with or scale up from the current quality. The rationale behind *user's desire for better quality* function $B(x_s, x_s^y)$ is described as follows. Intuitively, users are generally assumed to desire the best possible "value" for the money they pay. By this, it means a user may prefer to lower his/her requirement to achieve better perceived value. On the other hand, a user may demand more bandwidth for higher quality when

$$\beta_s \frac{x_s^{y+k} \lambda_s}{m_s} \le B(x_s, x_s^y), \quad (7)$$

where $\beta_s$ is a positive constant variable that indicates user's desire to save or spend money. Lower $\beta_s$ means *higher* willingness to spend more money for additional bandwidth to achieve better quality at $y+k$. Otherwise, the user may stay put with the current quality at layer $y$. Additionally, positive variable $k$ denotes the number of layers to be increased, $m_s$ denotes user's budget or willingness to pay for the service, and $\lambda_s$ is the network price. $(x_s^{y+k} \lambda_s)$ can be interpreted as the cost a user incurs for quality at layer $y+k$. Hence, the ratio $\frac{x_s^{y+k} \lambda_s}{m_s}$ shows that user's desire to for higher layer for better quality can be expressed through increasing $m_s$. Next, $B(x_s)$ is defined as follows.

$$B(x_s, x_s^y) = \frac{U_s^{y+k}(x_s^{y+k}) - U_s^y(x_s^y)}{x_s^{y+k}(\lambda_s + \lambda_s^{inc})}, \quad (8)$$

where $\lambda_s^{inc}$ denotes the *range of price a user is willing to increase* for the desired quality. By re-arranging eq. (7), the lower bound for the additional price range $\lambda_s^{inc}$ is

$$\lambda_s^{inc} \ge \left( \frac{m_s \left( U_s^{y+k}(x_s^{y+k}) - U_s^y(x_s^y) \right)}{\beta_s \lambda_s \left(x_s^{y+k}\right)^2} \right) - \lambda_s . \quad (9)$$

Observe in (9), the relationship between $\beta_s$ and $\lambda_s^{inc}$ is that lower $\beta_s$ (higher desire to spend) implies a greater range in the price a user is willing to pay. In other words, eq. (9) provides the lower bound for the additional cost a user must spend to achieve quality at layer $y+k$ at that given moment. Subsequently, given $x_s^y$, user may demand additional bandwidth to achieve higher quality if the new demand bandwidth $x_s^{y+k}$ satisfies (7). Thus,

$$x_s = \begin{cases} x_s^{y+k}, & \text{If condition (7) is } satisfied \\ x_s, & Otherwise \end{cases} . \quad (10)$$

That implies user may attempt to transmit data at $x_s^{y+k}$ instead of $x_s$, for $x_s < x_s^{y+k}$, when the condition allows.

*C. The impact of Adaptive User Demand*

According to the condition in (10), when (8) is not satisfied, then user must transmit data at $x_s$ and stop demanding additional bandwidth $x_s^{y+k}$.

***Proposition 1*** : When $x_s \ge x_s^y$, then $\frac{U_s(x_s, x_s^y)}{U_s^y(x_s^y)} = 1, \; as\, \lambda_s \to \infty$.

*Proof:* First, Observe that

$$B(x_s, x_s^y) = \lim_{\lambda_s \to \infty} \frac{U_s^{y+k}(x_s^{y+k}) - U_s^y(x_s^y)}{x_s^{y+k}(\lambda_s + \lambda_s^{inc})} = 0$$

and $\lim_{\lambda_s \to \infty} \beta_s \frac{x_s^{y+k} \lambda_s}{m_s} = \infty$, for $\beta_s > 0$. Since

$$\lim_{\lambda_s \to \infty} \sup \; \beta_s \frac{x_s^{y+k} \lambda_s}{m_s} = \infty ,$$

notice that $B(x_s) < \infty$ for any $\lambda_s$. Hence, as $\lambda_s \to \infty$, $\beta_s \frac{x_s^{y+k} \lambda_s}{m_s} > B(x_s)$. Consequently, when the condition in (8) is no longer satisfied, user $s$ stops demanding additional bandwidth beyond $x_s$. As a result, user $s$ settles with layer $y = \max(Y \mid x_s^y \le x_s)$, such that $\lim_{\lambda_s \to \infty} \sup x_s^y = x_s$. Hence, $\frac{U_s(x_s, x_s^y)}{U_s^y(x_s^y)} = 1$, as $\lambda_s \to \infty$. ■

Proposition 1 shows that users eventually stop demanding more bandwidth when additional quality is not worth the additional cost. Next, we investigate whether the algorithm with multi-layered user utility function also converges. First, we prove the following statement.

***Lemma 1:*** Suppose that $\lambda_l^*$ is an *optimal solution* for the dual problem $D(\lambda)$, where $\lambda_l^* > 0$ for link $\forall l,\; l \in L$, such that there exists a *subgradient* of $D(\lambda)$, $g(\lambda_l^*)$, at $\lambda_l^*$, where $g(\lambda_l^*) = 0$.

*Proof:* we have $\lambda_l^*$ as the minimizer of $D(\lambda)$, as $\forall l,\; l \in L$, there exists $g(\lambda_l^*)$ that satisfies

$$g(\lambda^*)^T \, |\lambda - \lambda^*| \ge 0, \qquad \text{for } \forall \lambda, \lambda \ge \bar{0}. \quad (11)$$

If we take $\lambda = \lambda'$, where $|\lambda_l' - \lambda_l^*| = \epsilon_l$, $\epsilon_l > 0$. By (11), we have $g_l(\lambda_l^*)\, \epsilon_l \ge 0$. Hence, when $\lambda_l^*$ is the optimal solution, $g_l(\lambda_l^*) = 0$. ■

We have shown that there exists subgradient $g_l(\lambda_l^*) = 0$, $\forall l,\; l \in L$, when $\lambda^*$ is an optimal solution for the dual problem $D(\lambda)$. Next, we investigate whether the algorithm converges with multi-layer user utility function.

***Proposition 2***: Suppose that $\lambda_l^*$ is an *optimal solution* for the dual problem $D(\lambda)$, where $\lambda_l^* > 0$ for link $\forall l,\; l \in L$. If $D(\lambda)$ is differentiable at $\lambda^*$, then $x(\lambda^t)$ converges $x(\lambda^*)$ as $\lambda^t$ converges to $\lambda^*$, for $t \to \infty$. Otherwise, $x(\lambda^t)$ and $\lambda^t$ may not converge.

*Proof:* Certainly, when dual problem $D(\lambda)$ is differentiable at $\lambda^*$, then $D(\lambda)$ has an unique subgradient at $\lambda^*$. Thus, $x(\lambda^*)$ is also unique. This means $x(\lambda)$ continues at $\lambda^*$, which implies that $x(\lambda^t)$ converges to $x(\lambda^*)$ and $\lambda^t$ converges to $\lambda^*$, for $t \to \infty$. By lemma 1, this includes when subgradient $g_l(\lambda_l^*) = 0$.

However, when $D(\lambda)$ is *not* differentiable at $\lambda^*$, then the subgradient at $\lambda^*$ is not unique. Thus, there exists a user with $x_s^{y_i+k}$, such that $x_s < x_s^{y+k}$. Given price at $\lambda^t$, by the condition in (10), $x_s(\lambda^t)$ is discontinued when condition (8) is *satisfied*. This implies the subgradient of $D(\lambda^t)$ at $\lambda^t$ is not unique. Thus, by eq. (10), $x_s(\lambda^t)$ may not converge. Furthermore, since $\lambda_l^{t+1}$ is a reflection of $\sum_{s\in S(l)} x_s(\lambda^t)$ in (6), $\lambda_l$ may not converge either. ■

***Lemma 2***: When $D(\lambda)$ is not differentiable at $\lambda^*$, there exists a link $l$ that satisfies this following condition:

$$\sum_{s\in l)} x_s(\lambda_l^t) < C_l \text{ and eq. (7) is satisfied.} \quad (12.\text{a})$$

$$\sum_{s\in l} x_s(\lambda_l^t) \geq C_l \text{ and eq. (7) is not satisfied.} \quad (12.\text{b})$$

*Proof:* When $D(\lambda)$ is not differentiable at $\lambda^t$, proposition 2 shows that there exists user $s$ with $\lambda_s^{inc}$ that satisfies condition (8). Thus, according to the condition in (10), $x_s(\lambda^t)$ discontinues at $\lambda^t$, then network ends up in condition (12.a). Furthermore, user $s$ may transmit data at rate $x_s^{y+k}$ at time $t$, for $x_s < x_s^{y+k}$. Since $\sum_{s\in l} x_s(\lambda_l^t) \geq C_l$, user must transmit at $x_s(\lambda^t)$. Thus, $x_s(\lambda^t)$ also discontinues at $\lambda^t$, then network ends up in (12.b). ■

Proposition 2 and lemma 2 imply that the rate allocation may oscillate between the two cases in (10) as a result from users attempting to obtain additional bandwidth for better QoS. The oscillation is also an indication there is a gap between the primal problem (4) and its dual problem $D$. The gap is driven by users' responses to the new prices advertised. That is, in one situation, users may feel the price is too high and decide not to demand additional bandwidth. In a different situation, the same users may demand additional bandwidth to achieve higher QoS when the price is acceptable to them. This makes solving the optimization problem with multi-layered user utility becomes intractable. Thus, there may be no optimal solution for the primal problem.

For this reason, we further investigate how this phenomenon may affect the network. Here, we divide users of multi-layered application into two categories: *Passive users* and *Active users*.

***Definition 1.*** *Passive Users:* Users who accept the amount of bandwidth $x_s$ allocated by the network and adjust the quality according to $x_s$, and achieve $x_s^y$ by solving $y = \max(Y \mid x_s^y \leq x_s)$.

***Definition 2.*** *Active Users:* Users who continue to try demanding additional bandwidth above the amount of bandwidth allocated to them.

These active users may cause oscillation as they change their transmission rate, which in turn affects the network pricing. They will stop demanding more bandwidth when they feel the additional quality is not worth the additional cost, or when the maximum quality is obtained. The question is therefore, how the behavior of active users affects network activities.

### D. *The Ripple Effects of Active Users on Network*

The following discussion addresses how the behavior of active users may affect the bandwidth allocation to passive users.

***Lemma 3****:* Suppose $\sum_{s\in S(l)} x_s(\lambda_l^t)$ on link $l$ oscillates, for $t \to \infty$, then $\lambda_l^t$ also oscillates.

*Proof:* Assume that $x_s(\lambda_l^t)$ oscillates as $t \to \infty$ and let $z_l^t = \sum_{s\in S(l)} x_s(\lambda_l^t)$ on link $l$. Since $\lambda_l^{t+1}$ is updated by eq. (6), for $\lambda_l^t \geq \lambda_l^{min} \geq 0$ and $z_l^{(t)} > 0$, when $z_l^t$ increases, then $\lambda_l^{t+1}$ also increases; and when $z_l^t$ decreases, then $\lambda_l^{t+1}$ also decreases. ■

Obviously, since eq. (6) is designed to respond to the traffic load in the network, the network price $\lambda_l^t$ oscillates when the traffic load oscillates. In fact, eq. (6) is a feedback loop and $\lambda_l^t$ continues to evolve as long as $\lambda_l^t \geq 0$. Hence, lemma 3 shows that active users can cause oscillation in pricing. Subsequently, we explore the effect of this pricing oscillation on bandwidth allocation for passive users.

***Proposition 3***: Bandwidth allocation for passive users is affected by the changes in network price caused by active users.

*Proof:* Let set $\hat{S}'(l)$ denote a set of *active* users and $\hat{S}(l)$ be a set of *passive* users sharing link $l$, where $\hat{S}(l) + \hat{S}'(l) = S(l)$, $\hat{s} \in \hat{S}(l)$, and $\hat{s}' \in \hat{S}'(l)$, for $l \in L$. Assume at time $t-1$, $x_{\hat{s}}(\lambda_l^{t-1})$ and $x_{\hat{s}'}(\lambda_l^{t-1})$ have converged at $\lambda_l^{t-1}$, for $\forall \hat{s}, \hat{s}'$ . Suppose at $t$, users in $\hat{S}'(l)$ demand more bandwidth and transmit data at $x_{\hat{s}'}^{y+k}$, where $x_{\hat{s}'}^{y+k} > x_{\hat{s}'}(\lambda_l^{t-1}) \geq x_{\hat{s}'}^{y}$, for $y = \max\left(Y \mid x_{\hat{s}'}^{y} \leq x_{\hat{s}'}(\lambda_l^{t-1})\right)$, but for users in $\hat{S}'(l)$, $x_{\hat{s}}(\lambda_l^t) = x_{\hat{s}}(\lambda_l^{t-1})$. Now, we have $\sum_{\hat{s}\in\hat{S}(l)} x_{\hat{s}}(\lambda_l^t) + \sum_{\hat{s}'\in\hat{S}'(l)} x_{\hat{s}'}^{y_i+k} > C_l$. Next, at $t+1$, $\lambda_l^{t+1}$ is updated by solving (6). Then, user $\hat{s} \in \hat{S}(l)$ computes $x_{\hat{s}}(\lambda_l^{t+1})$ by solving (5). By rearranging (2), we have

$$\lambda_{\hat{s}}^{t+1} = \frac{m_{\hat{s}}\left(U_{Bw}(x_{\hat{s}}) - U_{\hat{s}}(x_{\hat{s}}, x_{\hat{s}}^{y}) - 1\right)}{x_{\hat{s}}(\lambda_l)}.$$

$\lambda_{\hat{s}}^{t+1}$ is the network price that must be paid by user $\hat{s}$ from $U_{cost(\hat{s})}(x_{\hat{s}})$ in (2). Hence, the relationship between $\lambda_{\hat{s}}$ and $x_{\hat{s}}(\lambda_l)$ can be illustrated as follows.

$$\lim_{x_{\hat{s}}\to\infty} \lambda_{\hat{s}}(x_{\hat{s}}) = \lim_{x_{\hat{s}}\to\infty} \frac{m_{\hat{s}}\left(U_{Bw}(x_{\hat{s}}) - U_{\hat{s}}(x_{\hat{s}}, x_{\hat{s}}^{y}) - 1\right)}{x_{\hat{s}}} = 0.$$

However, we have

$$\lim_{x_{\hat{s}}\to 0} \lambda_{\hat{s}}(x_{\hat{s}}) = \lim_{x_{\hat{s}}\to 0} \frac{m_{\hat{s}}\left(U_{Bw}(x_{\hat{s}}) - U_{\hat{s}}(x_{\hat{s}}, x_{\hat{s}}^{y}) - 1\right)}{x_{\hat{s}}} = \infty.$$

Thus, when $\lambda_{\hat{s}(l)}^{t+1} < \lambda_{\hat{s}(l)}^{t}$, then $x_{\hat{s}}(\lambda_{\hat{s}(l)}^{t+1}) \geq x_{\hat{s}}(\lambda_{\hat{s}(l)}^{t})$. However, when $\lambda_{\hat{s}(l)}^{t+1} \geq \lambda_{\hat{s}(l)}^{t}$, then $x_{\hat{s}}(\lambda_{\hat{s}(l)}^{t+1}) < x_{\hat{s}}(\lambda_{\hat{s}(l)}^{t})$. ■

Proportion 3 implies that during excessive network congestion, oscillatory behavior exhibited by active users also impacts bandwidth allocation for passive users, which is consistent with the assumption that a heavier congestion leads to higher network price. As a result, passive users may end up with less bandwidth at a higher price. The *worst case scenario* is when the oscillation of network price $\lambda_l$ causes the bandwidth allocation for *passive user* $\hat{s}$ to oscillate between two cases: $x_{\hat{s}}(\lambda_l^t) \leq x_{\hat{s}}^{min}$ and $x_{\hat{s}}(\lambda_l^t) > x_{\hat{s}}^{min}$, where $x_{\hat{s}}^{min}$ is the minimum required bandwidth for minimum QoS. Therefore, the actions of active users may have a negative impact on the bandwidth allocation for passive users, such that passive users may not receive sufficient bandwidth even to meet the minimum requirement. Additionally, the oscillation also causes the quality of video to oscillate creating the visual flickering effect at the user level, that most people find annoying [27] and degrade user QoE. Therefore, we propose an admission control to assure QoE of users with multi-layered applications.

## V. ADMISSION CONTROL

In order to design an effective admission control (*adm ctrl*), network must decide the selection criteria to accept or reject users' requests for admission. In multi-layered user utility environment, user demand is adaptive and the long term consequence of poor QoS is potential loss of future revenue. We assume that admission control is invoked at the occurrence of excessive network congestion and each candidate for admission is evaluated with function $\vartheta_s(x_s(\lambda_s))$ defined as follows.

$$\vartheta_s(x_s(\lambda_s)) = x_s^y \left(\frac{\hat{\lambda}_s}{\lambda_s}\right)^{\delta_\lambda} + \delta_u \frac{U_s^y(x_s^y)}{\lambda_s\, x_s^y}, \qquad (13)$$

where $\delta_\lambda$ and $\delta_u$ are non-negative parameters that function as a weight: the increase in $\delta_\lambda$ implies that network puts more emphasis in revenue. Similarly, network places more weight in user utility when $\delta_u$ is increased.

Let $\lambda_s$ be the network price decided by the network and $\hat{\lambda}_s$ denote the *price user is willing to pay* for the desired service quality. That is $\hat{\lambda}_s = \lambda_s^{inc} + \lambda_s$, for $\lambda_s^{inc} \geq 0$. Observe that $\lim_{\delta_\lambda \to \infty} \left(\frac{\hat{\lambda}_s}{\lambda_s}\right)^{\delta_\lambda} = \infty$ when $\frac{\hat{\lambda}_s}{\lambda_s} > 1$. However, $\lim_{\delta_\lambda \to \infty} \left(\frac{\hat{\lambda}_s}{\lambda_s}\right)^{\delta_\lambda} = 0$, for $\frac{\hat{\lambda}_s}{\lambda_s} < 1$, as $\delta_\lambda \to \infty$. This means users with $\hat{\lambda}_s > \lambda_s$ receive higher preference for admission when network places more emphasis in revenue. Additionally, since user utility function with multi layers of quality follows the law of diminishing returns, network may consider $\frac{U_s^y(x_s^y)}{\lambda_s\, x_s^y}$ from eq. (13), which can be interpreted as user satisfaction over the cost for desired quality at layer $y$. Now we can formulate the admission control problem as the following optimization problem:

$$max \sum_{s \in S} \vartheta_s(x_s^y)\, z_s, \qquad (14)$$

$$s.t \sum_{s \in S(l)} x_s^y \leq C_l, \quad \text{for } \forall l, l \in L,$$

$$z_s \in \{0,1\}, \quad \text{for } \forall s, s \in S,$$

$$\text{over } x_s^y \geq 0, \quad \text{for } \forall s, s \in S \text{ and } y \in Y,$$

where $z_s = 1$ if user $s$ is selected, otherwise zero. The difficulty of solving this problem lies in the search for all possible combinations of $\vartheta_s(x_s^y)$, for $s \in S$. For this reason, problem (14) is reduced to the *0-1 Knapsack problem* [8], where each user must either be admitted or rejected. The network cannot admit a fraction of the amount of user's traffic flow or admit users above the available capacity. To ensure real-time performance and quick completion of the admission process, we propose a three-step heuristic greedy based algorithm to solve (14):

**Step one**: Network determines the price $\lambda_l$ of each link $l$. If $\lambda_l < \lambda_l^{min}$, then network sets the price $\lambda_l = \lambda_l^{min}$. This assures $\lambda_l \leq \lambda_l^{min}$. Next, network sends $\lambda_s$ to user $s$, where $\lambda_s = \sum_{l \in s} \lambda_l$

**Step two**: users submit a tuple of $\langle x_s^{y_i}, \hat{\lambda}_s \rangle$, where $\hat{\lambda}_s$ is the price user is willing to pay. Users respond to network after evaluating

$$x_s^y = \underset{0 \leq x_s^y \leq x_s^{|Y|}}{\arg\max} \left\{ \frac{U_s^y(x_s^y)}{x_s^y\, \lambda_s} \right\}.$$

**Step three**: Once network receive the necessary information, tuple $\langle x_s^y, \hat{\lambda}_s \rangle$, from the entire users, network computes $\vartheta_s(x_s(\lambda_s))$, $\forall s \in S$, and invoke "*User Selection*" algorithm.

**Algorithm 1:** *User Selection.*

1. $\vartheta_s^{\max} = \max\{\vartheta_{SET(l)}\}$
2. $x_s^y$ = get_bandwidth( $\vartheta_s^{\max}$ )
3. If $(x_s^y + \sum_{\hat{s} \in l} x_{\hat{s}} \leq C_l$, for $\forall l,\ l \in route\ r_s)$ and $(s \notin l)$ then
4. Reserve $l$ for user $s$, for $\forall l,\ l \in r_s$
5. $C_l = C_l - x_s^y$ for $\forall l,\ l \in r_s$
6. $\hat{S} = \hat{S} + s$
7. $\vartheta_{SET} = \vartheta_{SET} - \{\vartheta_s^{\max}\}$
8. Repeat from line 1 until $|\vartheta_{SET}| = 0$ *// until $\vartheta_{SET}$ is empty*

Let $\vartheta_{SET}$ denote a set of $\vartheta_s$ that is associated with user and $\hat{S}$ denote a set of users admitted into the network. In line 1 and 2 of *user selection* algorithm, given $\vartheta_s^{\max}$, $x_s^y$ is retrieved. In line 3, the algorithm verifies whether the link has sufficient capacity to provide at least $x_s^y$ and that $x_s^y$ has not been included from the previous run, and then, execute line 4, 5, and 6. Next, $\vartheta_s^{\max}$ is removed from the set in line 6. We assume that the network begins to provide service as soon as the user is admitted into network. The performance of this algorithm is determined by the number of links $|L|$, the number of users $|S|$, and the number of links in the path of each admitted user $s$ that need to be updated. Thus, the total running time is at most $O(|L|^2 . |S|)$.

## VI. SIMULATION AND DISCUSSION

In this section, we present a demonstration of multi-layered utility function with admission control using a network shared by eight users, shown in figure 2. The initial setup is listed in table 1 and table 2. Through this simulation, we demonstrate how user 3's switching between two layers may impact other users in the network and the pricing. We also show how implementing admission control improves network activities.

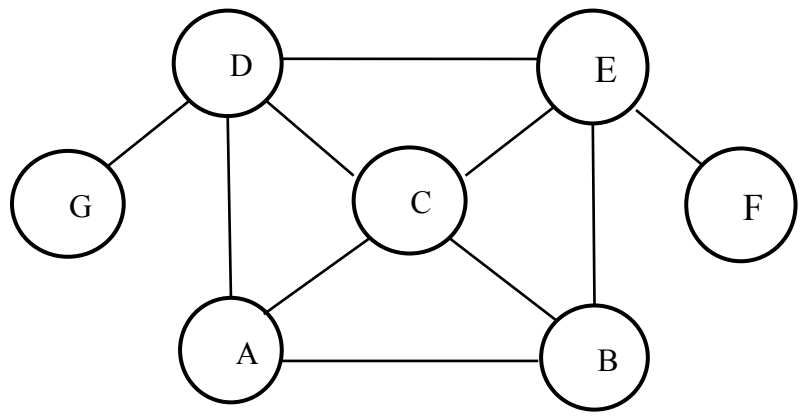

*Fig. 2. Network Topology.*

User 3 initially requests data transmission at 2, but demands bandwidth increase from 2 to 5 after iteration 70 when the network price is below his/her threshold of 6 unit currency, from solving eq. (8). User 3 would reverse his demand back to 2 when the network price exceeds the threshold. Next, network eventually performs an admission control at iteration 300 by solving problem (14) with Algorithm 1.

| User | 0 | 1 | 2 | 3 | 4 | 5 | 6 | 7 |
|---|---|---|---|---|---|---|---|---|
| $m_s$ | 40 | 50 | 50 | 50 | 70 | 10 | 100 | 70 |
| initial $x_s^{min}$ | 2 | 2 | 2 | 2 | 2 | 1 | 8 | 2 |
| New $x_s^{min}$ | - | - | - | 5 | - | - | - | - |

*Table 1. Simulation setup.*

| | | | |
|---|---|---|---|
| $U_0$: ABCD | $U_2$: ABC | $U_4$: DEF | $U_6$: DG |
| $U_1$: ABCDE | $U_3$: AB | $U_5$: CDG | $U_7$: EF |

*Table 2. User route or path setup.*

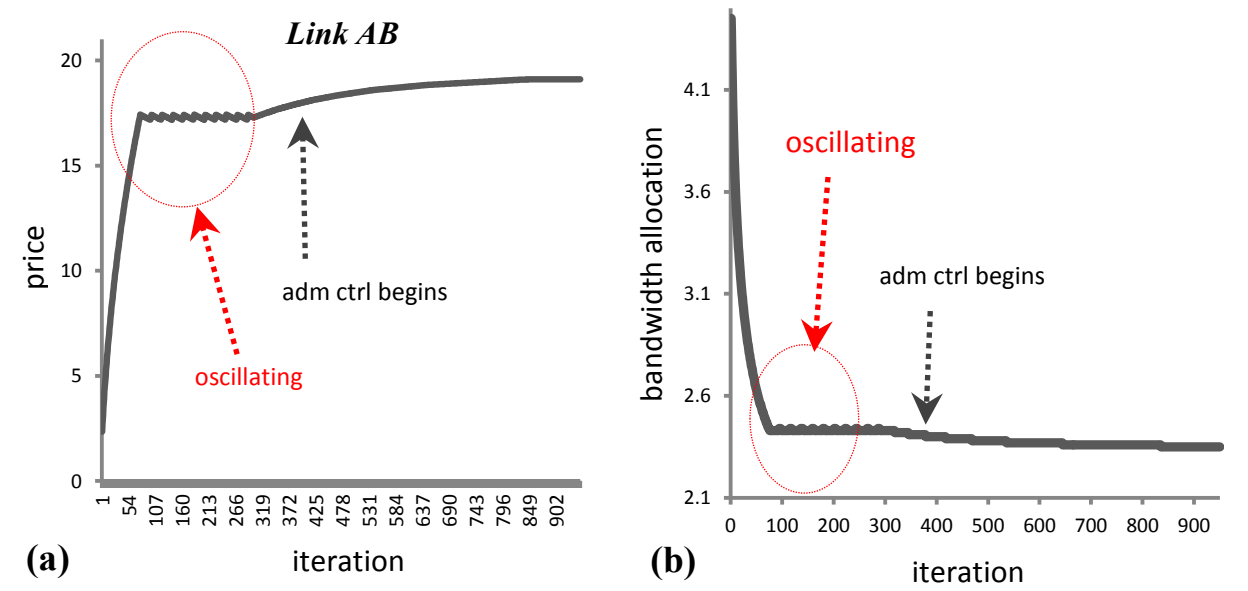

*Fig.3: Link price at link AB and bandwidth allocation of user 0,1, and 2. "**adm ctrl**" denotes admission control.*

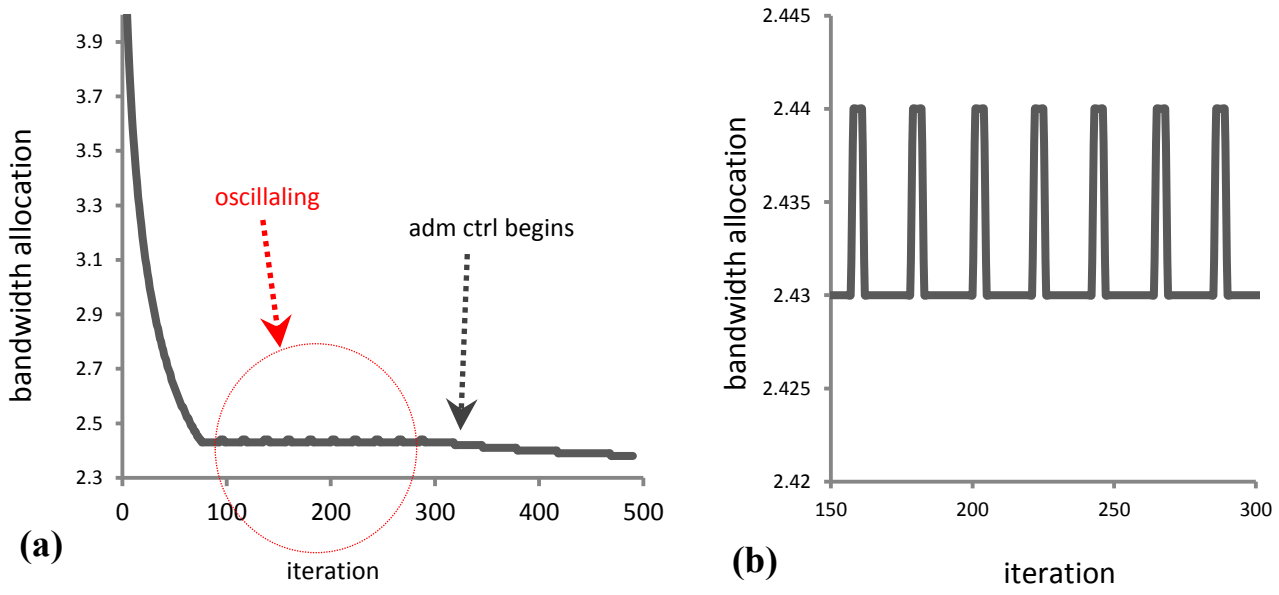

*Fig. 4. zoom into fig. 3.b between iteration 0-500 and iteration 150-300.*

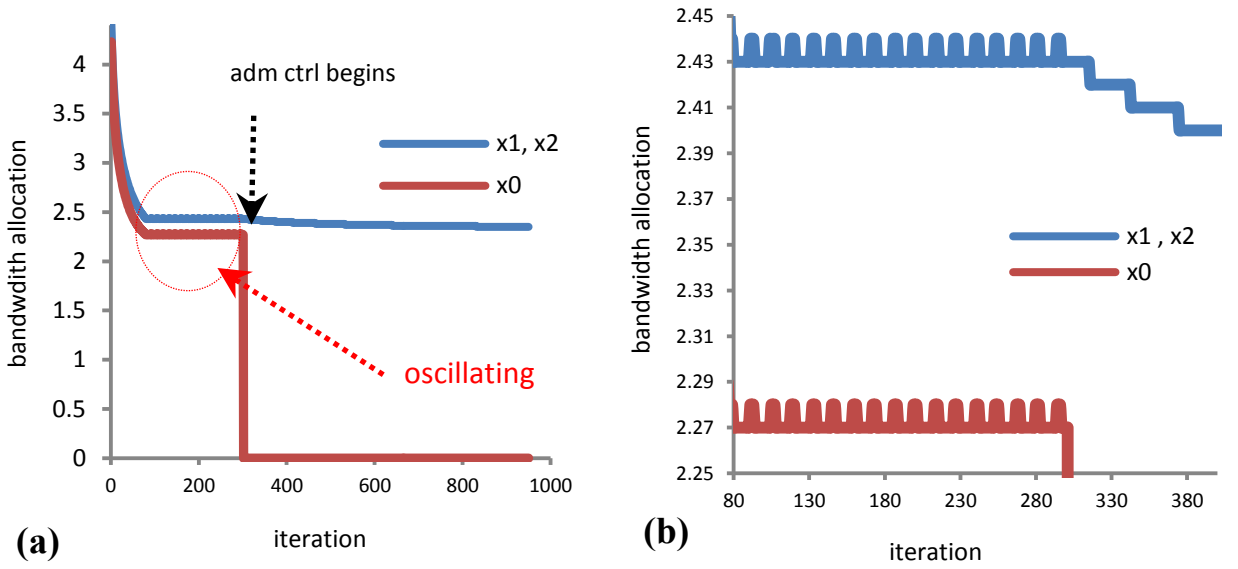

*Fig. 5. The impact of rate allocation to user 0, 1, and 2 in link AB. (Fig. 5.b magnifies the oscillating area in fig. 5.a.)*

We begin with our analysis of the most congested link, which is link AB that is shared by four users. These users are user 0, 1, 2, and 3. Prior to admission control, the network price (illustrated in figure 3.a) fluctuates whenever user 3 oscillates between two layers (figure 3.b). As a result, bandwidth allocation for user 0, 1, and 2 also oscillates, as illustrated in figure 5.a. At iteration 300, network implements admission control with *user selection* algorithm, resulting in the dismissal of user 0, and this in turn leads to the convergence of network price and bandwidth allocation for user 1, 2, and 3. In addition, without user 0, the network has additional bandwidth to meet the higher demand of user 3.

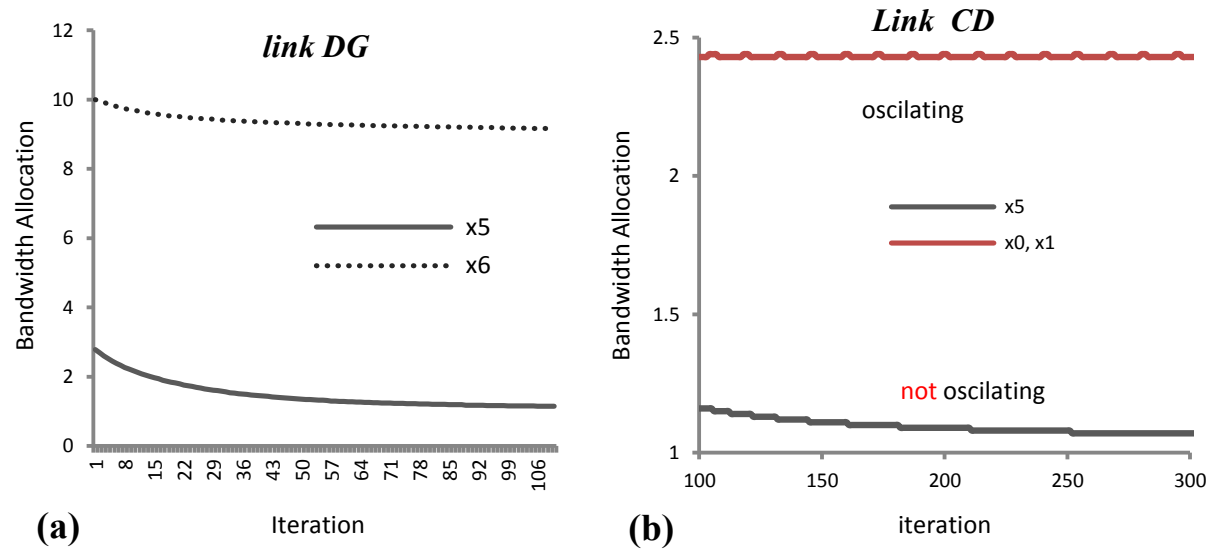

*Fig. 6. The rate allocation in link DG (fig. 6.a) and link CD (fig. 6.b).*

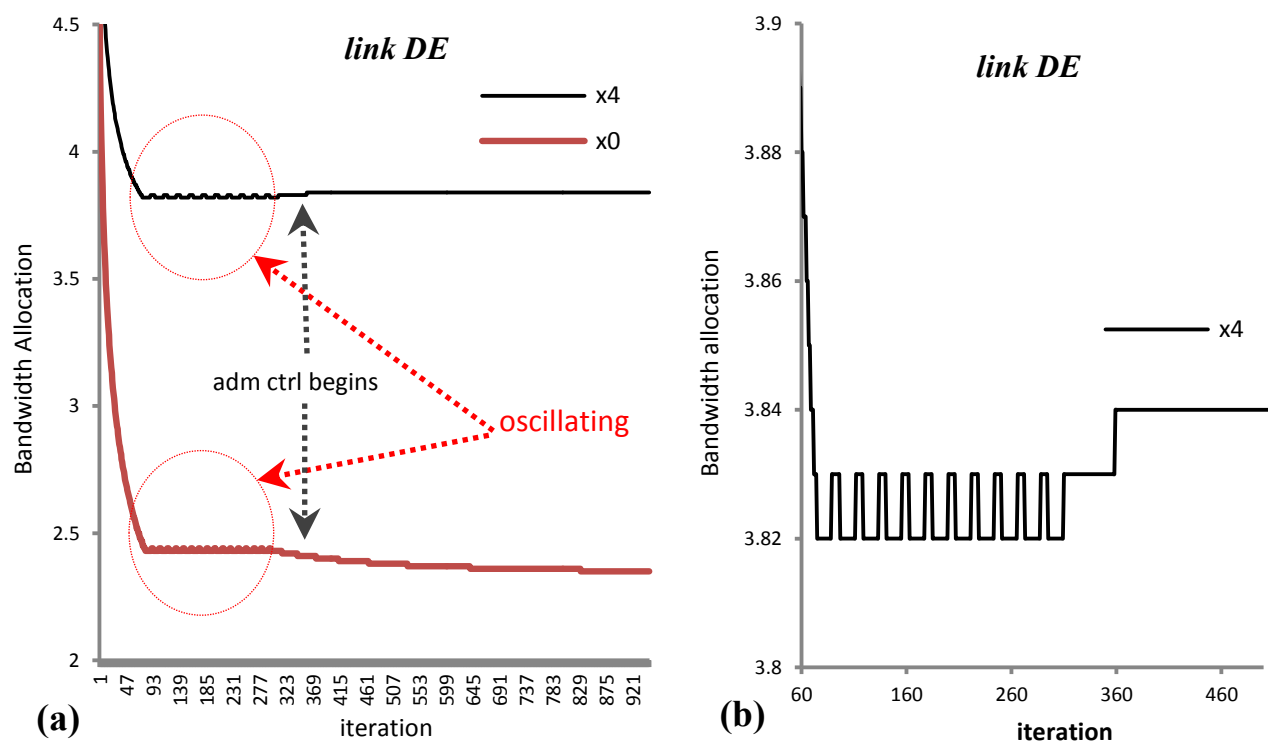

*Fig. 7. The rate allocation on link DE (fig. 7.a) and the oscillation area is maginify in fig. 7.b.*

Next, we examine the ripple effects from the oscillation caused by user 3. The rate allocation in link BC shows a pattern similar to the allocation in link AB because the same flows (user 0, 1, and 2) traversing through link BC also traverse in link AB. Figure 6.b shows that user 5 in link CD is *not* affected by the oscillation of user 0 and 1, which are reactions to the oscillation initiated by user 3. It is because the aggregated flow of the three users 0, 1 and 5, including the spike in rate allocation caused by the oscillation is still below the capacity constraint. Thus, the unused capacity in CD functions as a buffer for user 5, such that it allows user 0 and 1 to oscillate without affecting user 5. As demonstrated in figure 6.a, the rate allocation of user 5 and 6 in link DG converge and is not affected by the oscillation in CD. However, the oscillation from link AB is affecting user 4, such that his/her bandwidth allocation is also oscillating. This is because user 4 is sharing link DE with an oscillating flow that belongs to user 1, as illustrated in figure 7.a and 7.b. User 3's oscillation has impacted link DE because the aggregated flow exceeds the maximum capacity. Thus, any spike in the transmission rate causes congestion in DE. That is the aggregated flow in DE exceeds the capacity limit and it forces the network to hike the price at link DE. User 1 stops oscillating after the admission control is invoked at link AB, and this causes user 4 to also stabilize. Lastly, since user 7 is not sharing link EF with anyone, user 7 is not affected at all.

The simulation shows that active users striving for better QoE may cause many ripple effects, causing rate allocation assigned to other users to oscillate. Furthermore, the ripple effects of bandwidth oscillation may spread from one specific link to other parts of the network. The oscillation and its ripple effects confirm that solving optimization involving adaptive QoE makes the optimization problem become intractable. This is because there is a circular event of users continuously adjusting their transmission rate according to the fall and rise of the price. At the same time, price fluctuation follows and depends on the rise and fall in bandwidth demand. Additionally, the simulation also shows that higher throughput may result in higher risk of ripple effects, which lead to higher network instability. Such the ripple effects may cause more users experiencing the visual flickering effect and user QoE degradation. That is price and bandwidth allocation oscillation at different part of the network. Admission control by the network is a viable approach to attend to and stop oscillation. This is because reducing the population in the network provides sufficient bandwidth for the admitted users to increase their demand until the desired QoE is achieved. The lessons learned in overcoming the oscillation are: firstly, network may pick the higher value in the price oscillation as the network price, hence users may have to settle with the QoE they can afford. Secondly, network must assure that it has sufficient bandwidth for admitted users to be able to achieve the desired QoE.

## VII. CONCLUSION

General solutions for NUM problems in single-layered environment are not sufficient for traffic problems in multi-layered multimedia applications where user utility is adaptive. Our multi-layered user utility function incorporates insights from the fields of computer graphic and HCI, resulting in a user-utility function that considers human's natural visual ability to perceive changes in image quality, influencing their demand for desired QoE. This translates to a user utility demand that follows the law of diminishing returns. We demonstrate that the adaptive demand of some users cause oscillation that may ripple through the network, leading to lower aggregate QoE. Our study shows that optimization problem with users who constantly pursue better QoE makes the optimization problem intractable. This desire for better quality also causes the visual flickering effect at the video application level, which degrades user QoE. Thus, we propose a greedy based solution for admission control, such that the balance between revenue and user satisfaction can be achieved. This allows the rate allocation algorithm to converge. However, even when the network seeks to maximize its revenue during the process of admission control, the algorithm may not yield the maximum revenue because of the nature of greedy algorithm. Furthermore, the efficiency of multi-layer utility function must be investigated, which will be addressed as part of our future work.